\documentstyle[prc,aps,12pt]{revtex}
\def\be{\begin{equation}}
\def\ee{\end{equation}}
\def\bkk#1{<\kern-0.1167em \kern-0.1167em >} 
\begin{document}
\preprint{}
\title{Anharmonic collective excitation in a solvable model}
\author{G.F. Bertsch$^{1}$, P.F. Bortignon$^{2}$, and K. Hagino$^{1,3}$}
\address{$^{1}$Institute for Nuclear Theory, Department of Physics, 
University of Washington, \\ Seattle, WA 98195, USA \\
$^{2}$ Dipartimento di Fisica, Universit\`a di Milano \\ and 
INFN, Sezione di Milano, Via Celoria 16, I-20133 Milano, Italy \\
$^{3}$ Department of Physics, Tohoku University, Sendai 980--8578, Japan}
\maketitle

\bigskip

\begin{abstract}
We apply the time-dependent variational principle, the 
nuclear field theory, and the boson expansion method to 
the Lipkin model to discuss 
anharmonicities of collective vibrational excitations. 
It is shown that all of these approaches lead to the same
anharmonicity to leading order in the number of particles.
Comparison with the exact solution of the Lipkin model shows 
that these theories reproduce it quite well.

\end{abstract}   

\section{Introduction}

The existence of multiphonon states in the nuclear spectrum of excitation
has been predicted  since the introduction of collective
models \cite{BM75}. Examples of low-lying nuclear vibrational states have
been known for many years in nuclear spectra and are still being actively
investigated with new generation of detectors; in particular, two-phonon 
multiplets and some three-phonon states based on the low-lying 
collective quadrupole and
octupole modes have been found \cite{ABC87,CZ96,GVB90,KPZ96,YGM96,K92,POD97}.
Recently, it has also been beautifully demonstrated that multi-phonon 
excitations of these low-lying collective vibrations strongly influence 
heavy-ion fusion reactions at energies 
near and below the Coulomb barrier \cite{SACN95}, through the so-called 
fusion barrier distribution analysis\cite{RSS91,LDH95}. 
It was pointed out that anharmonicities of vibrational excitations 
can significantly alter the shape of fusion barrier distribution 
and that thus sub-barrier fusion reactions offer an alternative 
method for extracting  the static quadrupole moments of phonon 
states in spherical nuclei\cite{HKT98,HTK97}.

In the past 15 years, evidence has been collected for two-phonon
giant resonances as well\cite{ABE98}.
This evidence stems from heavy-ion reactions at
intermediate energy\cite{CF95,POL86}, pion-induced double charge exchange
reactions\cite{MO94}, and relativistic heavy-ion reactions via
electromagnetic excitations, in particular the excitation of the 
double giant dipole resonance (DGDR)\cite{E94}.
In the experiments of the last type, 
a puzzling problem has been reported\cite{SBE93,RBK93,AKS93,BSW96}. 
Although the experimental 
data show that the centroid of the DGDR is about 
twice the energy of the single phonon resonance, theoretical calculations 
which assume the harmonic oscillator for giant resonances 
considerably underestimate the cross sections for double phonon 
states. In connection with this problem, anharmonic properties of giant 
resonances are attracting much 
interests\cite{BF97,VCC95,LAC97,BD97,NW95,VBR95,AV92}. 

Recently Bertsch and Feldmeier applied the time-dependent variational 
principle (TDVP) \cite{KK76} to large amplitude collective motions and 
discussed anharmonicities of various giant resonances\cite{BF97}. One 
of the advantages of their approach 
is that solving the resultant equations 
and estimating the degree of anharmonicity are quite simple. 
They found that the relative frequency of the double phonon 
state scales as $\Delta\omega/\omega\sim A^{-4/3}$, 
$A$ being the atomic number of a nucleus. 

Earlier, 
Ponomarev et al. \cite{PBBV96} noted that this quantity scales as 
$A^{-2/3}$ in the nuclear field theory (NFT) \cite{BBDLSM76}, 
the same as the NFT result for the octupole mode \cite{BM75}. 
Reference \cite{BM75} also remarks that a liquid-drop estimate 
gives an A$^{-4/3}$ dependences, implying that quantal effects are 
responsible for the difference.  
Both NFT and TDVP are quantal theories giving different results
in refs. 
\cite{BBDLSM76} and \cite{BF97}, so differences must be due 
either to inadeqate approximations or differences in the underlying 
Hamiltonians. 
We therefore undertook to try both methods on a solvable
Hamiltonian. This will test the reliability of both methods, and if both
give correct results, the disagreement is very likely
attributable to the Hamiltonian assumptions.

The time-dependent variational approach was recently applied also to 
dipole plasmon resonances of metal clusters \cite{H99}. 
In Ref. \cite{H99}, it was shown that the anharmonicity of dipole 
plasmon resonances is very small and scales as 
$\Delta\omega/\omega\sim A^{-4/3}$, which is identical to 
the result of the time-dependent variational approach for nuclear 
giant resonances. 
On marked contrast, Catara {\it et al.} claimed a large anharmonicity 
of plasmon resonances using a boson expansion method\cite{CCG93}, 
which contradicts 
both to the result of Ref. \cite{H99} and to recent experimental 
observation \cite{SKIH98}. 
This fact also motivated us to compare these two methods, together 
with the nuclear field theory, on a solvable 
model in order to clarify the origin of the discrepancy. 

As an aside, we mention also that there is a size-dependent anharmonicity 
in quantum
electrodynamics, considering the photon spectrum in a small cavity.
In QED the only
dimensional quantity is the electron mass $m_e$, and the photon-photon
interaction is proportional to $m_e^{-4}$. Thus on dimensional grounds
the relative shift of a two photon state in a cavity of side $L$ scales
as $\Delta\omega/\omega\sim 1/m_e^4L^4$.
Considering that sizes of a system scale as $R\sim A^{1/3}$, 
the results of refs.
\cite{BF97,H99} are $\Delta\omega/\omega\sim R^{-4}$, identical to QED.  

In this work we compare the nuclear field theory, the time-dependent
variation approach, and the boson expansion method 
on the second Lipkin model, eq. (2.5) in ref.~\cite{LMG65}.
The model is finite-dimensional and can be solved exactly. 
It has widely been used in literature to test a number of 
many-body theories \cite{H73,FC72,BC92,KWR95,VCACL99}.

The paper is organized as follows. 
In Sec. II, we first solve the model in the random phase approximation (RPA) 
and discuss the harmonic limit. In Sect. III we derive the 
collective Hamiltonian using the TDVP.
We requantize the collective Hamiltonian and discuss the deviation from the 
harmonic limit. Numerical calculations are performed and compared with 
the exact solutions.  
In Sec. IV, we use the nuclear field theory as an alternative. There we
see that it gives the same result as the TDVP, to leading order
in the dependence on the number of particles.
In Sec. V, we compare those results with the boson expansion approach. 
It will be shown that it leads to the identical result to 
the TDVP and the NFT. Finally we summarise the paper in Sec. VI. 

\section{HARMONIC LIMIT}

Lipkin, Meshkov and Glick\cite{LMG65} proposed two Hamiltonian models to
describe $N$ particles, each of which can be in two states, 
making $2^N$ states in the basis. 
Using Pauli matrix representation
for the operators in the two-state space, the second model has the
Hamiltonian
\begin{equation}
\label{HH}
H=\frac{1}{2}\epsilon\sigma_z - \frac{V}{2}\sigma_x\sigma_x. 
\end{equation}
The first term is the single-particle Hamiltonian with an excitation energy
$\epsilon$, and the second term is a two-body interaction.
The quasi-spin operators $\sigma_z$ and $\sigma_x$ are given by 
\begin{eqnarray}
\sigma_z&=&\sum_{i=1}^{N}(a^{\dagger}_{1i}a_{1i}
-a^{\dagger}_{0i}a_{0i}),\\
\sigma_x&=&\sum_{i=1}^{N}(a^{\dagger}_{1i}a_{0i}
+a^{\dagger}_{0i}a_{1i}),
\end{eqnarray}
respectively. 
$a^{\dagger}_{1i}$ ($a_{1i}$) and $a^{\dagger}_{0i}$ ($a_{0i}$) 
are the creation (annihilation) operators of the 
$i$-th particle for the upper and the lower levels, respectively. 
For small $V$, the Hartree ground state $|0>$ is the fully 
spin-polarised state with matrix elements given by 
\begin{equation}
<0|\sigma_i|0> = -N\delta_{i,z}.
\end{equation}
A suitable basis for the exact diagonalization of $H$ is the set of
eigenvectors
of the angular momentum operators $J^2$ and $J_{z}$ with $J=N/2$.  
Then the
dimension of the matrix diagonalization is reduced from $2^N$ to
$N+1$, making the numerical problem very easy.

Before going to the anharmonicity, we note that the harmonic limit
is obtained by solving the RPA equations. This was carried out in ref. 
\cite{MGL65} for the first Lipkin model. 
The RPA frequency for the second Lipkin model, Eq. (\ref{HH}), 
is obtained in the same manner. 
Setting the RPA excitation operator $O^{\dagger}$ as 
\begin{equation}
O^{\dagger}=
X\left(\sum_{i=1}^{N}a^{\dagger}_{1i}a_{0i}\right) -Y
\left(\sum_{i=1}^{N}a^{\dagger}_{0i}a_{1i}\right),
\label{RPAexc}
\end{equation}
the RPA equations read
\begin{equation}
\left(\begin{array}{cc}
A&B\\
B&A
\end{array}\right)
\left(\begin{array}{c}
X\\
Y
\end{array}\right)
=\omega
\left(\begin{array}{cc}
1&0\\
0&-1
\end{array}\right)
\left(\begin{array}{c}
X\\Y
\end{array}\right),
\end{equation}
where $A$ and $B$ are given by
\begin{eqnarray}
A&=&\frac{1}{4N}<0|[\sigma_-,[H,\sigma_+]]|0>
= \epsilon (1-\chi), \\
B&=&-\frac{1}{4N}<0|[\sigma_-,[H,\sigma_-]]|0>
= \epsilon \chi, 
\end{eqnarray}
respectively. We have defined $\sigma_-$, $\sigma_+$, and $\chi$ as 
\begin{eqnarray}
\sigma_-&=&2\sum_{i=1}^{N}a^{\dagger}_{0i}a_{1i}, \\
\sigma_+&=&2\sum_{i=1}^{N}a^{\dagger}_{1i}a_{0i}, \\
\chi&=&V(N-1)/\epsilon,
\end{eqnarray}
respectively. 
From these equations, the RPA frequency and the amplitude of the forward 
and the backward scatterings are found to be
\begin{eqnarray}
\label{RPA}
\omega&=&\sqrt{(A+B)(A-B)}=\epsilon\sqrt{1-2\chi}, \\
X&=&\frac{\omega+\epsilon}{2\sqrt{N\omega\epsilon}}, \\
Y&=&\frac{\omega-\epsilon}{2\sqrt{N\omega\epsilon}}, 
\end{eqnarray}
respectively. 

Fig.~1 compares the exact solution for the excitation energy of the
first excited state with the RPA frequency given by eq.~(\ref{RPA}).
As a typical example, we set the strength of the interaction 
$V/\epsilon$ to be 0.01.
The solid line shows the exact solutions for this particular choice 
of the Hamiltonian parameters, while the dashed line shows the RPA 
frequency. At $N=51,~ \chi=0.5$, the RPA frequency becomes zero and 
the system undergoes phase transition from spherical 
to deformed. 

Figure 2 is the same as Fig.~1, but for a fixed $\chi$.
We set $\chi$ to be 0.25, which corresponds to the isoscalar giant 
quadrupole resonance. 
We find significant deviation of the RPA frequency from the exact 
solution for small values of $N$, suggesting large anharmonicities. 
We discuss now the deviation from the harmonic limit. 

\section{TIME-DEPENDENT VARIATIONAL APPROACH}

The time-dependent variational approach has been applied  to the first
Lipkin model by Kan et al. in ref. \cite{KA80},
but has never been applied to our knowledge to the second model. 
In keeping with the procedure of ref. \cite{BF97}, we postulate
a time-dependent wave function of the form
\be
\label{ab}
|\alpha\beta\rangle = \exp(i\alpha(t) \sigma_x)\exp(i\beta(t)
\sigma_y)|0\rangle.
\ee
The motivation for this ansatz appears in ref.~\cite{BF97}.  The operator
in the first term is the one that we wish to evaluate in the 
transition matrix elements.  The operator in the second term
is obtained by the commutator with the Hamiltonian,
\be
[H,\sigma_x] = i \epsilon \sigma_y.
\ee

The Lagrangian is given by 
\be
\label{L}
{\cal L} = -{\dot \alpha} \langle \beta|\sigma_x|\beta\rangle - \langle
\alpha\beta|H|\alpha\beta\rangle.
\ee
We reduce this with the help of the identity
\be
e^{-i\sigma_i \theta} \sigma_j e^{i\sigma_i \theta}= \cos 2\theta\, \sigma_j
+ \sin 2\theta\, \sigma \cdot (\hat i \times \hat j),
\ee
where $i\ne j$ are Cartesian indices of the Pauli matrices.  For example,
the bracket in the first term of eq.~(\ref{L}) is reduced as
\be
\langle \beta|\sigma_x|\beta\rangle = \langle 0|
e^{-i\sigma_y \beta} \sigma_x e^{i\sigma_y \beta}|0\rangle
= \cos 2\beta \langle 0|\sigma_x|0\rangle - 
\sin2\beta \langle 0|\sigma_z|0\rangle.
\ee

The first term in the Lagrangian is 
\be
-{\dot \alpha} \langle \beta|\sigma_x|\beta\rangle 
= -N\dot\alpha\sin 2\beta.
\ee
The second term is
\be
- \langle\alpha\beta|H|\alpha\beta\rangle=
\epsilon {N\over 2}\cos 2\alpha\,\cos
2\beta\,  +V{N\over 2}(\cos^2 2\beta +N\sin^2 2\beta\, ).
\ee
The Lagrangian may then be expressed as
\be
\label{LL}
{\cal L} = -N{\dot \alpha} \sin2\beta +\epsilon{N\over 2} \cos 2\alpha\,\cos
2\beta\,  +V{N\over 2} (\cos^2 2\beta +N\sin^2 2\beta\, ).
\ee
The first Lagrangian equation is $d/dt\,\partial{\cal L}/\partial\dot\alpha
-\partial {\cal L}/\partial\alpha=0$.  It reduces to
\be
\dot\beta = {\epsilon\over 2} \sin 2 \alpha.
\ee
Similarly from the second Lagrange equation, 
$d/dt\,\partial{\cal L}/\partial\dot
\beta-\partial {\cal L}/\partial\beta=0$ we obtain 
\be
\dot\alpha\cos2\beta + {\epsilon\over 2} \cos 2\alpha\, \sin 2\beta
-(N-1)V \cos2\beta\,\sin 2\beta=0.
\ee
Next let us linearize and see what the RPA frequencies would be.  The
linearized equations are
\be
\dot\beta = \epsilon \alpha
\ee
$$
\dot\alpha + (\epsilon - 2V (N-1))\beta=0.
$$
The equation for
the frequency reads
\be
\omega^2=\epsilon^2 - 2 (N-1) V \epsilon =\epsilon^2(1-2\chi)
\ee 
in agreement with the result of eq.~(\ref{RPA}).

A Hamiltonian corresponding to our Lagrangian can be
seen by inspection, comparing eq.~(\ref{LL}) to the form
\be
{\cal L} = \dot q  p - {\cal H}(p,q).
\ee
Equation~(\ref{LL}) is already of this form with e.g., 
$p= - {N\over 2}\sin 2 \beta$, $q= 2\alpha$. The Hamiltonian is
then given by
\be
{\cal H}(p,q) = -\epsilon {N\over 2} \cos q \sqrt{1-(2p/N)^2} - V{N\over 2}
\Bigl((1-(2p/N)^2) + N (2p/N)^2\Bigr).
\label{HTDV}
\ee
We now expand $\cal H$ in powers of $q$ and $p$ up to fourth order.  
Dropping the constant term, the expansion has the form
\be
\label{H}
{\cal H} = {p^2\over 2 m} + {k\over 2} q^2 + a q^4 + b p^4 + c q^2 p^2
\ee   
with coefficients
\begin{eqnarray}
{1\over m} &=& {2\over N}\Biggl(\epsilon-2 V (N -1)\Biggr), \\
k &=& \epsilon {N\over 2}, \\
a &=& -{\epsilon N \over 48}, \\
b &=& {\epsilon\over N^3}, \\
c &=& -{\epsilon\over 2N}.
\label{c}
\end{eqnarray}
Note that we recover the linear frequency, eq.~(\ref{RPA}), immediately from
$\omega^2 = k/m$.  

In ref. \cite{BF97}, the anharmonicity was determined  by requantizing
the Hamiltonian with the Bohr-Sommerfeld condition,
\be
\int_{q0}^{q1} p dq = n \pi,
\ee
where $p$ and $q$ satisfy ${\cal H}(p,q)=E$ and $q_0$ and $q_1$ are 
the endpoints of the motion at energy $E$.  
However, here we find it more convenient to use the
equivalent formula
\be
\label{BS}
\int_{{\cal H} < E} dp dq = 2n\pi.
\ee
In the same sense as the expansion of the Hamiltonian ${\cal H}(p,q)$ as 
done in eq. (\ref{H}), we apply eq.~(\ref{BS}) iteratively. 
We first consider only the harmonic part of the 
Hamiltonian and transform the integration region to a circle.
We also use polar coordinates and write $p'=p/\sqrt{m}=r \sin\theta,
xq=\sqrt{k}q =\cos\theta$.  The radius of the circle is then 
$r_0=\sqrt{2E}$ and the harmonic approximation gives
\be
\int_{{\cal H} < E} dp dq = \sqrt{m\over k} \int_{{\cal H} < E} dp' dq'
= {1\over \omega} \int_0^{2\pi}  d\theta \int_0^{r_0} r dr
= 2\pi {E\over\omega}.
\ee
The nonlinearity can now be treated as a perturbation.  To lowest
order in the quartic terms, the radius to the boundary surface is
given by
\be
r_1^2\approx r_0^2 - 2 r_0^4 \Biggl({a\over k^2} \cos^4\theta+
m^2 b \sin^4\theta + {m c\over k}\sin^2\theta\cos^2\theta\Biggr).
\ee
This integral is easily evaluated with the result
\be
{1\over 2 \pi}\int_{{\cal H} < E} dp dq \approx  {E\over\omega} -
{\omega\over 2}\left({E\over\omega}\right)^2
\left(3{a\over k^2}+3{b m^2}+{c m\over k}\right).
\ee
Inserting the parameters from eqs.~(\ref{H})-(\ref{c}), the anharmonic
term can be expressed
\be
{\omega\over 8\epsilon N} \left({E\over\omega}\right)^2
\left(-1 + 3(\epsilon/\omega)^4- 2(\epsilon/\omega)^2\right).
\ee
Note that if there is no interaction, $\omega=\epsilon$ and
the anharmonicity vanishes.  This is rather remarkable; the
Hamiltonian in this case is the first term in eq. (\ref{HTDV}), which
looks nonlinear.  But the solution of the equations of motion
are independent on excitation energy.  It is not a harmonic oscillator
spectrum, however, because the energy is bounded.  These two
properties correspond exactly to the quantum spectrum of the
operator $\epsilon J_z$.

We next quantize the above action to get
\be
E_n = n\omega + n^2{\omega^2\over 8\epsilon N} 
\Bigl(-1 + 3(\epsilon/ \omega)^4- 2(\epsilon/\omega)^2\Bigr).
\ee
Taking the second difference, this yields an anharmonicity of
\be
\label{ANH}
\Delta^{(2)} E ={\omega^2\over 4\epsilon N} \Bigl(
-1 + 3(\epsilon/\omega)^4- 2(\epsilon/\omega)^2\Bigr) 
=\frac{2\chi\epsilon^3}{N\omega^2}\left(1-\frac{\chi}{2}\right).
\ee

The exact value of anharmonicity  $\Delta^{(2)}E$ is compared with the
value obtained from eq.~(\ref{ANH})  in Fig. 3. We can see that the
time-dependent variational principle works very well.

\section{Nuclear Field Theory (NFT) Approach}

The NFT is a formulation of many-body perturbation theory with 
vibrational modes summed to all orders in RPA. 
Its building blocks are RPA phonons 
and the single particle degrees of freedom 
which are described in the Hartree-Fock approximation. 
The coupling between them is treated diagramatically in the 
perturbation theory. 
For the Hamiltonian (\ref{HH}), the effective NFT Hamiltonian 
is given, to the lowest order, by 
\begin{equation}
H_{NFT}=\frac{1}{2}\epsilon\sigma_z + \omega O^{\dagger}O 
+H_{pv}. 
\label{HNFT}
\end{equation}
The first term in the $H_{NFT}$ describes single-particle spectrum. In 
writing down this term, we have used the fact 
that, for small value of $V$, the excitation energy 
in the HF is given by $\epsilon$ and that the creation and the annihilation 
operators for the HF levels are the same as those for unperturbative levels. 
The second term describes the RPA phonons, with 
$\omega$ and $O^{\dagger}$ given 
by eqs. (\ref{RPA}) and (\ref{RPAexc}), respectively. 
The particle-vibration interaction $H_{pv}$ in eq. (\ref{HNFT}) is 
given as
\begin{equation}
H_{pv}=-\Lambda (O^{\dagger}+O)\sigma_x,
\end{equation}
where the coupling constant $\Lambda$ is given by 
\begin{equation}
\Lambda=NV\sqrt{\frac{\epsilon}{N\omega}}=\epsilon\chi'
\sqrt{\frac{\epsilon}{N\omega}},
\end{equation}
$\chi'$ being $NV/\epsilon$. 
This Hamiltonian is constructed by replacing $\sigma_x$ in the 
two-body interaction in the original Hamiltonian (\ref{HH}) 
as in Ref. \cite{BBDLS76}. 
In general, there is also a residual interaction among 
particles and holes, but it 
does not contribute for the Hamiltonian (\ref{HH}) to the lowest order. 

Each graph in the NFT contributes to a given order in $1/N$,
but to all orders in $\chi'$. 
Since the microscopic origin of the RPA phonon is a coherent sum of 
particle-hole excitations, bubble diagrams have to be excluded 
when one calculates physical quantities in the NFT \cite{BBDLSM76}. 
To the zero-order (in $1/N$), 
the phonon energy coincides with that in the RPA given by eq. (\ref{RPA}). 
The anharmonicity begins with the leading $1/N$ diagrams,
which are shown in Fig. 4. 
These diagrams are called \lq\lq butterfly\rq\rq graphs (see also 
refs. \cite{BM75,BBDLSM76,HA74}). 
For each diagram shown in fig.4, there are 5 other diagrams which are 
obtained by changing the direction of the phonon lines. 
As already said, for the Hamiltonian (\ref{HH}) there is no diagram of order 
$1/N$ involving residual interaction among fermions.

The contribution from each diagram is most easily evaluated by 
using the Rayleigh-Schr\"odinger energy denominator, which are more 
suitable in the lowest order expansion \cite{BBB78}.
The four graphs in Fig. 4 have identical contributions, each given by 
\be
Graph(a)={-N \Lambda^4\over{(2\omega -\omega 
-\epsilon )^2(2\omega-2\epsilon)}}.
\ee
In this equation, the minus sign appears 
because of the crossing of two fermion lines \cite{BBBL77}.
By summing up the contributions from all diagrams, we obtain 
\be
\Delta^{(2)}E=
-4N\Lambda^4\frac{\epsilon(\omega^2+3\epsilon^2)}{(\omega^2-\epsilon^2)^3}
=\frac{2\chi'\epsilon^3}{N\omega^2}\left(1-\frac{\chi'}{2}\right). 
\label{ANHNFT}
\ee
To compare with eq.~(\ref{ANH}), note that 
$\chi'=\chi$ to the leading order of $1/N$. 
With this substitution, the two results are identical.

\section{BOSON EXPANSION APPROACH}

In the boson expansion method, each fermionic operators is 
replaced by corresponding operators which are written in terms 
of boson operators. There are several prescriptions to carry 
out the mapping from the fermionic to the bosonic spaces \cite{RS81}. 
Here we follow Refs. \cite{BC92,VCACL99,PKD68} which discussed 
the anharmonicities of the Lipkin model using the Holstein-Primakoff 
mapping. 
In this mapping, fermionic operators are mapped to bosonic 
operators so that the commutation relations among operators 
are preserved. 
The quasi-spin operators in the present two-level problem are then 
mapped as \cite{BC92,VCACL99,PKD68,RS81} 
\begin{eqnarray}
\sigma_+ &\to & 2\sqrt{N}B^{\dagger}\sqrt{1-\frac{B^{\dagger}B}{N}} \\
\sigma_- &\to & 2\sqrt{N}\sqrt{1-\frac{B^{\dagger}B}{N}} B\\
\sigma_z &\to & -N + 2 B^{\dagger}B, 
\end{eqnarray}
where the operators $B$ and $B^{\dagger}$ satisfy the boson commutation 
relation, i.e., $[B, B^{\dagger}]=1$. 
The Hamiltonian in the boson space which corresponds to 
the Lipkin Hamiltonian (\ref{HH}) is therefore obtained as 
\begin{eqnarray}
H_B &\sim& \epsilon\left(1-\frac{NV}{\epsilon}\right)B^{\dagger}B
-\frac{NV}{2}\left(B^{\dagger}B^{\dagger}+BB\right) \nonumber \\
&&
+VB^{\dagger}B+\frac{V}{4}\left(B^{\dagger}B^{\dagger}+BB\right) 
+VB^{\dagger}B^{\dagger}BB 
+\frac{V}{2}\left(B^{\dagger}B^{\dagger}B^{\dagger}B+B^{\dagger}BBB\right), 
\label{BH}
\end{eqnarray}
to the first order in $1/N$. 

A truncation of the expansion up to the leading order of the $1/N$ 
corresponds to the RPA which we discussed in Sec. II. 
To this order, the boson Hamiltonian (\ref{BH}) is given by 
\begin{equation}
H^{(2)}_B=
\epsilon\left(1-\frac{NV}{\epsilon}\right)B^{\dagger}B
-\frac{NV}{2}\left(B^{\dagger}B^{\dagger}+BB\right). 
\label{BH2}
\end{equation}
As is well known, 
this Hamiltonian can be diagonalised by introducing a transformation 
\begin{eqnarray}
B^{\dagger}&=& X_0O^{\dagger} + Y_0O \label{RPAOD} \\
B&=& X_0O + Y_0O^{\dagger}, \label{RPAO} 
\end{eqnarray}
and imposing 
\begin{equation}
[H^{(2)}_B, O^{\dagger}]=\omega O^{\dagger}, 
\end{equation}
with a condition $X_0^2-Y_0^2=1$. 
The frequency $\omega$ then reads 
\begin{equation}
\omega=\epsilon\sqrt{1-2\chi'},
\end{equation}
$\chi'$ being $NV/\epsilon$, which was introduced in the 
previous section, together with 
\begin{eqnarray}
X_0^2&=&\frac{\omega+\epsilon(1-\chi')}{2\omega}, \\
Y_0^2&=&\frac{-\omega+\epsilon(1-\chi')}{2\omega}, \\
X_0Y_0&=&\frac{\epsilon \chi'}{2\omega}.
\end{eqnarray}
The frequency $\omega$ coincides with that in the RPA given by 
eq. (\ref{RPA}) to the leading order of $1/N$. 

As in the nuclear field theory approach, the anharmonicity 
begins with the next order of the $1/N$ expansion. 
In terms of RPA phonon creation and annihilation operators defined 
in eqs. (\ref{RPAOD}) and (\ref{RPAO}), 
the boson Hamiltonian (\ref{BH}) can be rewritten as 
\begin{eqnarray}
H_B&=&\omega O^{\dagger}O+H_{11}O^{\dagger}O
+H_{20}\left(O^{\dagger 2}+O^2\right) \nonumber \\
&&+H_{40}\left(O^{\dagger 4}+O^4\right) 
+H_{31}\left(O^{\dagger 3}O+O^{\dagger}O^3\right) 
+H_{22}O^{\dagger 2}O^2, 
\end{eqnarray}
where the first term is the leading order of the $1/N$ expansion 
given by eq. (\ref{BH2}) and the rest are the higer order corrections.  
The coefficient $H_{22}$, for example, is given by 
\begin{equation}
H_{22}=V(X_0^4+3X_0^3Y_0+4X_0^2Y_0^2+3X_0Y_0^3+Y_0^4)
=\frac{V\epsilon^2}{\omega^2}\left(1-\frac{\chi'}{2}\right).
\end{equation}
In order to estimate the degree of anharmonicity, we use 
the perturbation theory. The first order perturbation gives 
the energy of the one and the two phonon states of 
\begin{eqnarray}
E_1&=&\omega+H_{11}, \\
E_2&=&2(\omega+H_{11}) + 2H_{22},
\end{eqnarray}
respectively. Taking the second difference, the anharmonicity 
reads
\begin{equation}
\Delta^{(2)}E=2H_{22}=
\frac{2\chi'\epsilon^3}{N\omega^2}\left(1-\frac{\chi'}{2}\right),
\end{equation}
which is identical to the result of the variational approach given by eq. 
(\ref{ANH}) as well as that of the nuclear field theory, eq. (\ref{ANHNFT}). 

\section{Conclusion}

We have shown that the nuclear field theory, the time-dependent
variational principle, and the 
boson expansion method give identical leading-order anharmonicities for the
Lipkin model, and that the formulas agree well with the exact numerical 
solution.
The anharmonicity is inversely proportional to the number of particles in the
system, when the other parameters are fixed to keep the harmonic frequency
the same.
This clarifies the origin of the conflicting results for the $A$-dependence
of the anharmonicity obtained in ref. \cite{BF97} and \cite{BBDLSM76}.
In ref. \cite{BF97} the time-dependent method was applied to a Skyrme-like
Hamiltonian involving all $A$ nucleons, and the result was $\Delta^{(2)}E
\propto f(\omega)/A$. In ref. \cite{BBDLSM76}, the Hamiltonian
was restricted to a space of a single major shell for particle orbitals
and similarly for the hole orbitals.
Since the number of particles in the valence shell increases as
$A^{2/3}$, the result was $\Delta^{(2)}E\propto f(\omega)/A^{2/3}$.
Finally, it should perhaps be emphasized that both methods predict that the
anharmonicity is very small for giant resonances: both $A^{2/3}$ and $A$
are large numbers. This need not be the case for low-lying collective
vibrations. The NFT can produce large effects when there are small energy
denominators. Low-lying excitations are difficult to describe with a simple
ansatz like eq.~(\ref{ab}), so the time-dependent variational principle is
not easily applied. Clearly this is an area that should be explored further.

As for the discrepancy between the time-dependent variational approach 
and the boson expansion method concerning anharmonicities of 
plasmon resonances of metal clusters, our study showed that 
the origin of the discrepancy should not 
be ascribed to the method used to solve the problem. 
The origin of the discrepancy, therefore, is not traceable at 
moment and further studies 
may be necessary in order to reconcile the discrepancy. 

\section*{Acknowledgments}
We thank A. Bulgac for showing us the Hamiltonian construction and 
C. Volpe for discussions. 
K.H. acknowledges support from the Japan Society for the Promotion of
Science for Young Scientists.  
G.F.B. acknowledges support from the U.S.
Dept. of Energy under Grant DE-FG-06ER46561. 
Both G.F.B. and P.F.B. thank
the IPN of Orsay for warm hospitality where this work was completed.

\newpage

\begin{center}
{\bf Figure Captions}
\end{center}

\noindent
{\bf Fig. 1:}
Energy of the one phonon state as a function of the number of particle. 
The solid line is the exact solution of the Lipkin model, whilist 
the dashed line is obtained in the RPA. 
The strength of the interaction $V/\epsilon$ is set to be 0.01. 

\noindent
{\bf Fig. 2:}
Same as fig. 1, but for a fixed $\chi$ parameter, which is set to be 
0.25. 

\noindent
{\bf Fig. 3:}
Anharmonicity of the double phonon state as a function of the 
number of particle for $\chi=0.25$. 
The dashed line is an approximate solution given by eq. (\ref{ANH}), while 
the solid line is obtained by numerically diagonalising the Hamiltonian. 

\noindent
{\bf Fig. 4:}
The diagrams of order $1/N$ contributing to the anharmonicity of the 
double phonon state for the Hamiltonian (\ref{HH}). The wavy lines represent 
phonon propagations, while the particles and the holes are 
depicted by the allowed lines. 

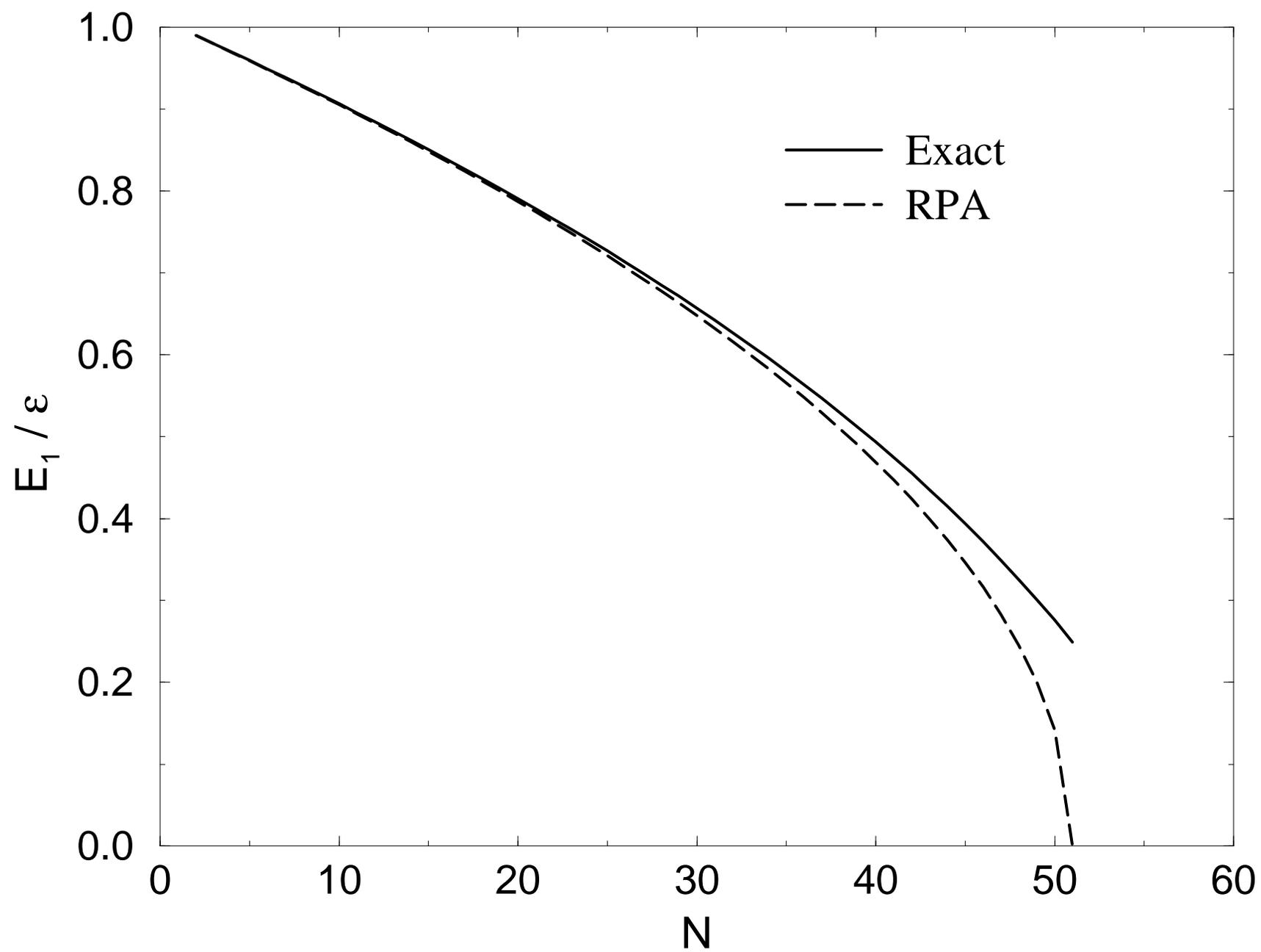
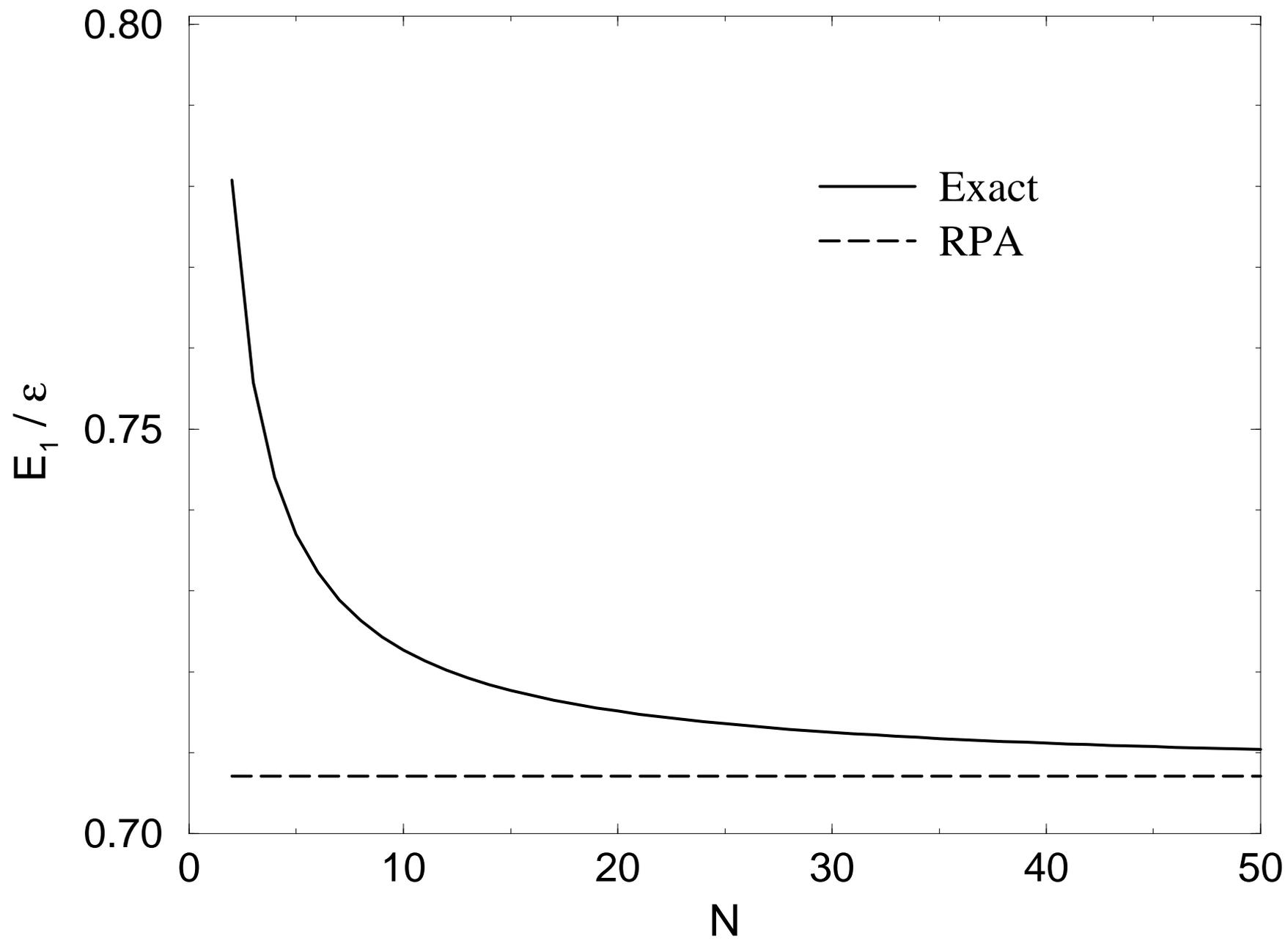
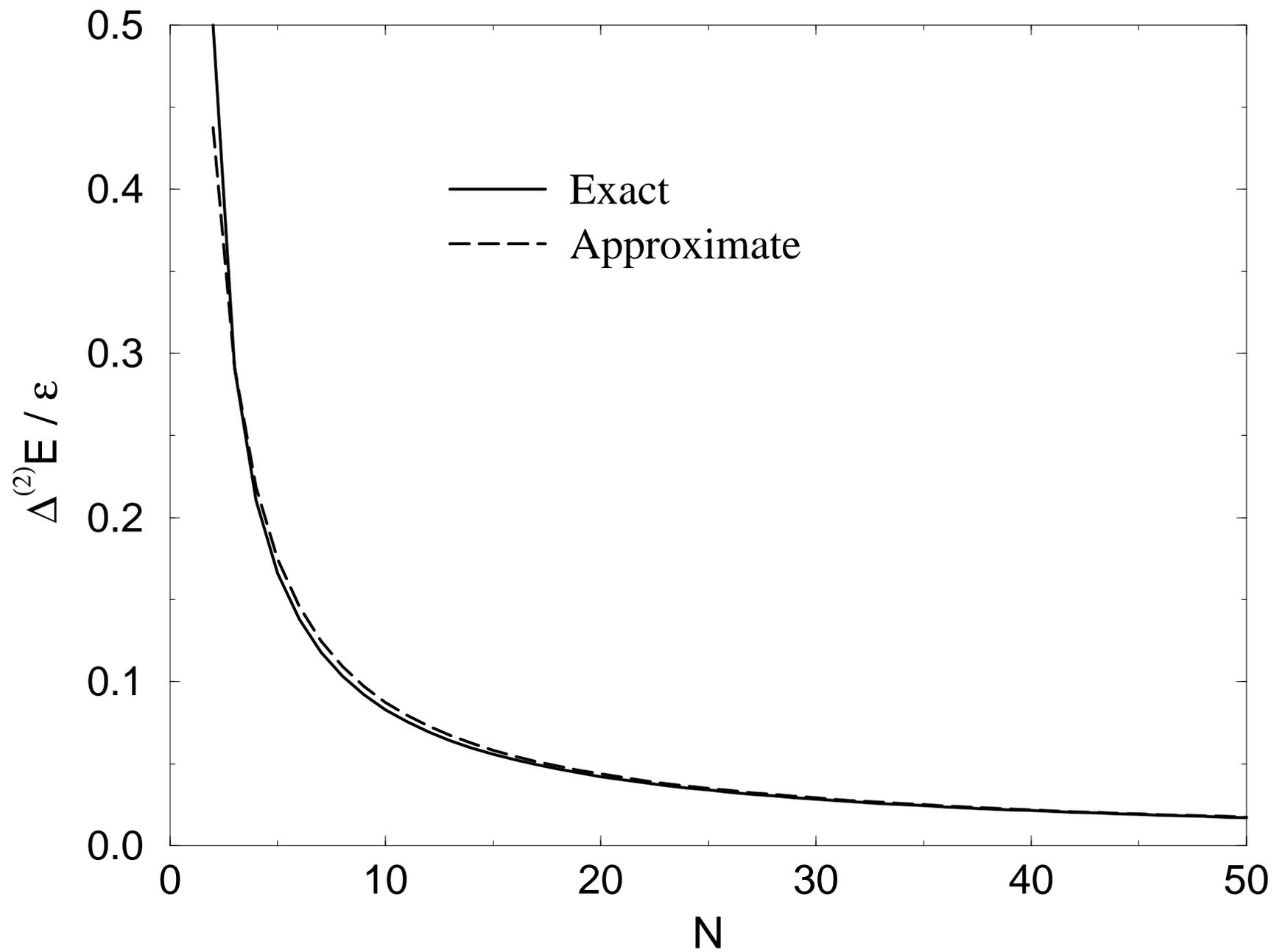
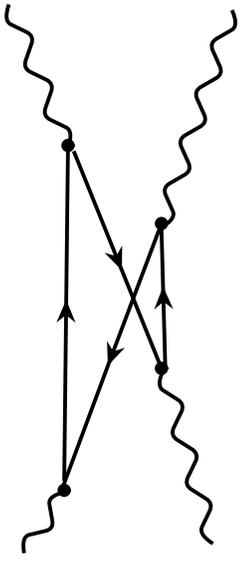
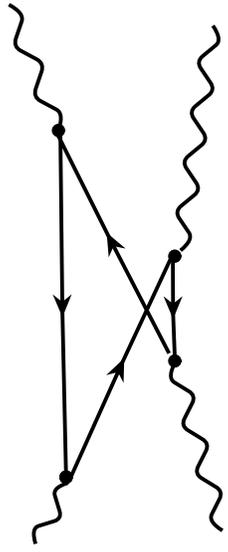
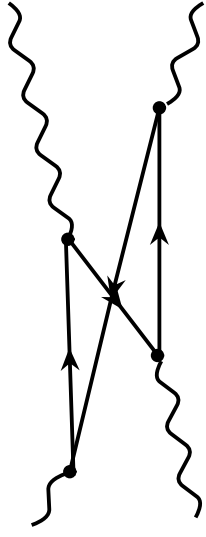
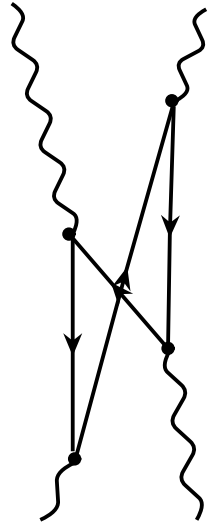
\end{document}